\documentclass[prb,twocolumn,grouppedaddress,nopacs]{revtex4}
\usepackage{times}
\usepackage{amsfonts}
\usepackage{amsmath}
\usepackage{graphicx}
\usepackage{subfigure}
\usepackage{wasysym}

\begin{document}
\bibliographystyle{apsrev}


\title{Dzyaloshinskii--Moriya interactions in valence bond systems}

\author{Mayra~Tovar}
\affiliation{Department of Physics and Astronomy, University of California at Riverside, Riverside,
 CA 92507}
\author{Kumar~S.~Raman}
\affiliation{Department of Physics and Astronomy, University of California at Riverside, Riverside,
 CA 92507}
\author{Kirill~Shtengel}
\affiliation{Department of Physics and Astronomy, University of California at Riverside, Riverside,
 CA 92507}

\date{\today}

\begin{abstract}

We investigate the effect of Dzyaloshinskii--Moriya interactions on the low temperature magnetic susceptibility for a system whose low energy physics is dominated by short-range valence bonds (singlets).  Our general perturbative approach is applied to specific models expected to be in this class, including the Shastry--Sutherland model of the spin-dimer compound SrCu$_2$(BO$_3$)$_2$ and the antiferromagnetic Heisenberg model of the recently discovered $S=1/2$ kagom\'{e} compound ZnCu$_3$(OH)$_6$Cl$_2$.  The central result is that a short-ranged valence bond phase, when perturbed with Dzyaloshinskii--Moriya interactions, will remain time-reversal symmetric in the absence of a magnetic field but the susceptibility will be nonzero in the $T\rightarrow 0$ limit.  Applied to ZnCu$_3$(OH)$_6$Cl$_2$, this model provides an avenue for reconciling experimental results, such as the lack of magnetic order and lack of any sign of a spin gap, with known theoretical facts about the kagom\'{e} Heisenberg antiferromagnet.

\end{abstract}


\maketitle


\section{Introduction}

A question at the heart of frustrated magnetism is what happens to an
antiferromagnet when conventional magnetic states, such as the N\'{e}el state, are
destabilized by frustrating interactions or geometry.  Classically, a system
commonly finds itself in a degenerate manifold of configurations
which share a local constraint, and whose size grows exponentially with the
system\cite{Wegner71}.  For quantum spins, this would violate the third law of
thermodynamics.  Instead, it has been suggested that the system will enter one
of a number of exotic non-magnetic phases\cite{Anderson73, Read89b, Wen91b},
with orders that do not have classical analogs.  While some of these exotic
quantum phases have now been seen in toy models\cite{Rokhsar88,Misguich99,
Moessner01a, Balents02,Misguich02,Raman05}, settling the question of their
existence as a matter of principle, an unambiguous experimental observation of
such phases in actual materials is still lacking.  Therefore, the issue of what experimental
signatures are indicative or at least suggestive of such phases is a question
of considerable interest.  The most basic feature is the \emph{absence} of
magnetic order even when the temperature is much lower than the
antiferromagnetic coupling $J$.

Within a non-magnetic phase, one may visualize the system wave function as a
superposition of \emph{valence bond states}, i.\ e.\ states
where every spin is paired with another spin to form a singlet or
\emph{valence bond}.  The valence bond states form a highly overcomplete
basis for the $S=0$ subspace\cite{Beach06} so the representation need not be unique.
However, it is convenient to distinguish between phases in which the wave function \emph{can} be expressed mainly in terms of \emph{short-range} valence bond states\cite{Kivelson87},
where the singlet pairing is always between \emph{nearby} spins, and phases
where the description always involves important contributions from valence bonds of all
lengths.  In the latter case, the most famous example of which is the
original formulation of the resonating valence bond
(RVB) liquid \cite{Anderson73,Anderson87}, the spin--spin correlation decays
algebraically and the magnetic excitations are gapless.  In the former case,
which includes the valence bond solids\cite{Read89b} and the short-range
$Z_2$ and $U(1)$ RVB
liquids\cite{Moessner01a, Huse03}, spin--spin correlations decay
exponentially.
While, in principle, the short-ranged nature of equal time correlations does
not necessarily imply a spectral gap \cite{FNS05} (and hence it
might be possible to conceive of a phase with exponential spin--spin
correlations and
gapless magnetic excitations), the most familiar examples of
short-range
valence bond phases, including the three just mentioned,
have a spin gap.  The intuition for this is that
an elementary magnetic excitation can be viewed as ``breaking" a valence
bond by replacing it with a triplet (or a pair of $S=1/2$ ``spinons"),
which costs an energy of order $J$ if the bond is between nearby spins but a
vanishingly small amount if the bond is very long.  An experimental
consequence of having a spin gap $\Delta$ is that the magnetic susceptibility
should vanish exponentially at low temperatures: $\chi\sim e^{-\Delta/T}$.

These issues have gained additional prominence in light of recent experiments
probing
the magnetic properties of the recently discovered
compound ZnCu$_3$(OH)$_6$Cl$_2$, also known as
herbertsmithite.  In this material, the magnetic properties are determined by
the $S=1/2$ copper ions which arrange themselves in nearly perfect, widely
separated, kagom\'{e} planes \cite{Shores05}.  Measurements of
the powder magnetic susceptibility at high temperatures, when fitted to a
Curie--Weiss
law, reveal an antiferromagnetic exchange constant $J \simeq 200$~K
\cite{Helton07}.  A
variety of different measurements \cite{Helton07,Ofer07,Mendels07,Imai08}
confirm that the material shows no evidence of magnetic order down to
temperatures as low as 50~mK $\sim 10^{-4} J$, which suggests a non-magnetic
ground state.

The simplest model consistent with these facts is the two-dimensional $S=1/2$
kagom\'{e} Heisenberg antiferromagnet (KHAF).  Indeed, while less is known
about the KHAF than its counterparts on other lattices, such as the
triangular,\footnote{The frustration of the classical triangular Heisenberg
antiferromagnet is partially relieved by having the spins arrange in a
non-collinear 120 degree pattern which turns out to closely model the ground
state of the quantum model\cite{Bernu92}, despite initial thoughts that it was
a spin
liquid\cite{Anderson73}.  In contrast, for the classical kagom\'{e} Heisenberg
model, the
ground state is highly degenerate\cite{Singh92} and there is no obvious reason
we are aware
of for why quantum fluctuations should favor one of them, though a particular
set of
them are apparently favored classically due to ``order by
disorder''\cite{Chalker92}.} it is widely
believed that its ground state is non-magnetic\cite{Leung93}.  Exact
diagonalization studies indicate a small spin gap\cite{Waldtmann98}
$\Delta\sim J/20$
and below this scale, the spectrum shows a large number of singlet states, the
number
growing exponentially with system size\cite{Mila98}.  These facts suggest a
picture
where the low energy physics of the KHAF, and hence ZnCu$_3$(OH)$_6$Cl$_2$, is
dominated by short-range valence bonds and valence bond solids
are, in fact, among the proposed ground states.\cite{Marston91,Nikolic03,Singh07a}

However, the coherence of this perspective is disturbed by the
puzzling fact that the material shows no sign of a spin gap
and the powder magnetic susceptibility does not go to zero as $T\rightarrow 0$.
While the original experiments\cite{Helton07}, in fact,
showed the susceptibility continuing to \emph{increase} even at temperatures small
compared to $J/20$, subsequent experiments have suggested that the susceptibility
might eventually saturate\cite{Ofer07} or perhaps decrease before eventually saturating
at a nonzero value.\cite{Olariu08}

A number of proposals have been
made to resolve this discrepancy including suggestions that the KHAF ground state may actually
be a gapless liquid state involving long-range valence bonds\cite{Ran07a};
the gaplessness is a disorder effect involving magnetic defects\cite{Bert07} and/or
nonmagnetic impurities\cite{Bert07, Gregor08a};
the true spin Hamiltonian of the material is closer to an Ising
model\cite{Ofer08};
and that Dzyaloshinskii--Moriya (DM) interactions play an important
role\cite{Rigol07a, Rigol07b}.
In this paper, we focus on this last idea but will comment on the other
suggestions further below.

\begin{figure}[t]
\includegraphics[width=0.9\columnwidth]{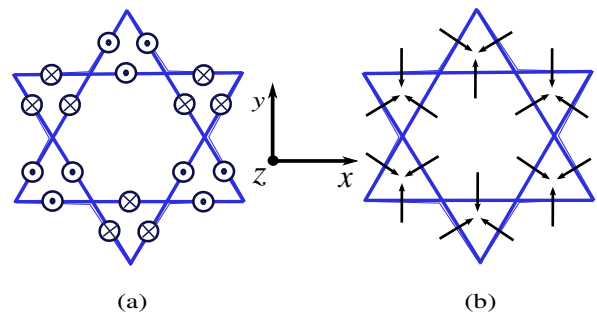}
\caption{The relations between the directions of the (a) out-of-plane and (b) in-plane
components of the DM vectors for nearest-neighbor interactions in the kagom\'{e} lattice are summarized.  The directions are determined by the physical requirement that $H_\text{DM}$ preserves the symmetries of the lattice, but also by the convention of how we
orient the links as they appear in the Hamiltonian.  The figure uses the convention of Ref.~[\onlinecite{Rigol07b}] where the DM interaction on the link $\langle ij \rangle$ appears in the Hamiltonian with the lower spin or, on horizontal links, the left spin, being the first member of the cross product.  A different choice, which is sometimes convenient, is where the links are oriented clockwise around each triangle. \cite{Elhajal02} In this convention, all of the $D^z$'s come with the same sign.}
\label{fig:DMdirecs}
\end{figure}

Microscopically, the DM interaction between spins arises due to the
spin--orbit coupling. Originally proposed on the basis of symmetry
\cite{Dzyaloshinskii57,Dzyaloshinsky58}, it can be derived microscopically as
a linear (in the spin-orbit coupling) correction to the
standard superexchange mechanism\cite{Moriya60}. The interaction has the form:
\begin{equation}
{H}_{\text{DM}}=\sum_{\langle ij\rangle} {H}_{\text{DM}}^{\langle ij \rangle}
=\sum_{\langle ij\rangle}\mathbf{D}_{ij}
\cdot (\mathbf{S}_i \times \mathbf{S}_j)
\label{eq:DM}
\end{equation}
where the sum is over pairs of spins on a lattice.  As with the Heisenberg
interaction, the dominant contribution comes from the
pairs of nearest neighbors, to which the sum is
commonly restricted.  Because the interaction is antisymmetric, we also need
to choose a convention for how the pair $\langle ij\rangle$ is oriented, i.e. whether it
appears in the Hamiltonian as $\mathbf{S}_i
\times \mathbf{S}_j$ or as $\mathbf{S}_j \times \mathbf{S}_i$
(see Fig.~\ref{fig:DMdirecs}).  Eq.~(\ref{eq:DM}) can be
viewed more formally as the antisymmetric part of the most general bilinear
interaction between spins.
\footnote{In fact, a careful microscopic derivation shows that
the a \emph{full} bilinear interaction between the spins is akin to the
familiar Heisenberg interaction: ${H}=\sum J \mathbf{S}_i \cdot
\mathbf{S}'_j$, where $\mathbf{S}'$ has been rotated about $\mathbf{D}_{ij}$
by a certain angle\cite{Kaplan83,Shekhtman92,Shekhtman93}. While
experimentally confirmed\cite{Zheludev98}, and undoubtfully important in
discussions of a spontaneous ferromagnetism, this observation is not essential
for the lowest order calculations of the uniform susceptibility in the absence
of magnetic order -- the subject of this manuscript.}

The vectors $\{ \mathbf{D}_{ij} \}$ are constrained by crystal symmetries to follow certain rules.  One rule is that if spins $i$ and $j$ both lie in a mirror plane of
the lattice, then $\mathbf{D}_{ij}$ must be perpendicular to this plane.  For a literally two-dimensional crystal embedded in three-dimensions, the lattice itself is a mirror plane so only the out-of-plane component $D^z$ will be present.  However, in
ZnCu$_3$(OH)$_6$Cl$_2$, this symmetry is broken by the (OH) groups which mediate the superexchange between Cu ions, so the in-plane component $\mathbf{D}^\text{p}$ will also be present.
Another rule is that $\mathbf{D}_{ij} =0$ if the midpoint of the line connecting spins
$i$ and $j$ is a center of inversion.  For the square and triangular lattices, the
midpoint of every such line is a center of inversion so for perfect lattices, DM will
not be present. However, the midpoints of the bonds forming a kagom\'{e} lattice are not centers of inversion\cite{Elhajal02} so a DM interaction between nearest-neighbor spins
can exist.  The relations between different $\mathbf{D}_{ij}$'s are
determined by the requirement that $\mathbf{D}_{ij}$ transforms like a vector
under a symmetry operation.  For the kagom\'{e} lattice, these relations are summarized in Fig.~\ref{fig:DMdirecs}.

The suggestion of DM coupling as a means of reconciling the
presence of a spin gap in the KHAF with its apparent absence in the real
material was first explored by Rigol and Singh in
Refs.~[\onlinecite{Rigol07a,Rigol07b}].  The authors noted
that the DM coupling would mix the singlet and triplet sectors so total spin is no
longer a good quantum number.  Instead, the ground state will contain $S=1$
components which contribute to a nonzero susceptibility at $T=0$.
The authors studied the KHAF augmented by the nearest-neighbor DM interaction given by Eq.~(\ref{eq:DM}). Fitting their results to the experiment, they estimated the values of $|\mathbf{D}^{\text{p}}|$ and $|D^z|$
to be $\sim 0.2 J - 0.3 J$
and $\sim 0.1 J$
respectively.  However, they also pointed
out that the
numerical techniques used in their study were most reliable at temperatures
larger than $0.3 J$.
On the other hand, one may argue that the observed susceptibility behaves
truly ``anomalously" only at temperatures smaller than the putative spin gap, since only at
these temperatures do we expect $\chi$ to start decreasing
due to singlet formation.  For the KHAF, such an energy scale is $\sim {J}/{20}$.\cite{Waldtmann98}  Therefore, as Rigol and
Singh noted in their
conclusion, a theory taking into account DM interactions in the low temperature regime is much needed.

In this paper, we are attempting to fill this gap
by studying ${H}_{\text{DM}}$ at the $T=0$ limit, which
emphasizes the quantum nature of the system.  Our main conclusion is that the picture of
ZnCu$_3$(OH)$_6$Cl$_2$ having a low temperature phase dominated by
short-range valence bonds can be reconciled with the susceptibility measurements if we
include the effect of a perturbatively small DM interaction.

Our approach is not specific to the kagom\'{e} lattice and we begin in section
\ref{sec:model}
by discussing an analytical method for calculating the susceptibility of a
valence bond
system in the presence of weak DM interactions on an arbitrary lattice
provided that the unperturbed Hamiltonian
satisfies certain assumptions.  The assumptions we require are: (1) the
unperturbed Hamiltonian has a narrow band of low energy short-range valence bond states
separated from magnetic states by a spin gap, (2) that $H_\text{DM}$ connects
this narrow $S=0$ band to a narrow band of $S=1$ magnetic states,
and (3) that the unperturbed Hamiltonian conserves the total spin.
For such models, we show that the DM interaction leads to a nonzero value
of the zero temperature susceptibility.

Then, in section \ref{sec:ss}, we apply our approach to
a model of the spin-dimer compound SrCu$_2$(BO$_3$)$_2$, which lives
on a Shastry--Sutherland lattice.  The reason for considering this model first
is that the assumptions and approximations of section \ref{sec:model}
hold exactly there.  Then, in section \ref{sec:check}, we consider a generalized Klein
model
on the checkerboard lattice \cite{Raman05, Nussinov07a}.  In this case,
we have a Hamiltonian whose ground states are known to be short-range
valence bond states;  whose spectrum has approximately the form we assume;
and for which the overlap expansion is believed to converge fairly well.  This
will
provide a level of confirmation that our methods can yield a correct order of
magnitude in a ``real'' problem.  In section \ref{sec:kagome}, we
return to the kagom\'{e} lattice and  ZnCu$_3$(OH)$_6$Cl$_2$.
In section \ref{sec:discuss}, we discuss our results in light of
recent numerical work, other experiments, and
alternative theoretical viewpoints.

\section{Model and Formalism}
\label{sec:model}

In this section, we calculate the effect of a small DM interaction on the
low temperature powder susceptibility of a lattice spin system whose low
energy properties are determined by short-range valence bonds.
Our calculation applies for a system that satisfies a few rather general
assumptions, which will be stated at the beginning of
section ~\ref{sec:perturbation}.  For simplicity, we concentrate
on the case where the valence bonds are always between nearest neighbor spins
but the argument is more general.  We begin by discussing some
technical aspects of working in a (nearest-neighbor) valence bond basis.

\subsection{Dimer basis and overlap expansions}

\begin{figure}[t]
\centering
\includegraphics[width=0.9\columnwidth]{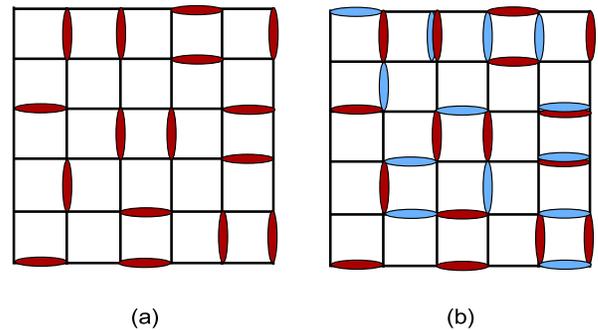}
\caption{(Color online) (a) A general dimer covering of the square lattice.  This represents
a wave function, which we write as $|d_\alpha\rangle$, where a dimer across a link, shown as a
thick red (black) line, means
that the two spins are in a singlet bond. (b) Overlaying two different dimer coverings, shown by the
red and blue (black and gray) lines, gives the transition graph, which will contain both double bonds, where the two coverings coincide, and closed loops of various even lengths.}
\label{fig:trangraph}
\end{figure}

A configuration where every spin is in a singlet with one of its neighbors can
be represented pictorially as a dimer covering of the lattice, as shown in Fig.~\ref{fig:trangraph}a
for a square lattice.  Therefore, we
will refer to these basic configurations $\{ |d_\alpha\rangle \}$ as ``dimer coverings''.
The wave functions which will
interest us can be written as superpositions of these dimer coverings:
\begin{equation}
|\phi \rangle=\sum_\alpha a_{\alpha} |d_\alpha\rangle
\label{eq:super}
\end{equation}
but this requires some care because the dimer coverings are not orthogonal.
One
consequence is that the inner product of $|\phi\rangle$ with a state
$|\psi\rangle=\sum_\alpha b_\alpha | d_\alpha\rangle$ is not simply
$\sum_\alpha a_\alpha^{*} b_\alpha$ but:
\begin{equation}
\langle \phi | \psi\rangle = 1 = \sum_{\alpha,\beta} a_{\alpha}^{*} b_{\beta}
\Omega_{\alpha\beta}
\end{equation}
where $\Omega_{\alpha\beta}\equiv \langle d_{\alpha}|d_\beta\rangle$ is the
overlap matrix.

The magnitude of the overlap of two dimer coverings is most easily calculated
by overlaying the configurations to form their \emph{transition graph}, as shown in
Fig.~\ref{fig:trangraph}b.
As shown in the figure, the resulting picture contains double bonds and loops
of various even
lengths.  The magnitude of the corresponding overlap matrix element is:
\begin{equation}
|\Omega_{\alpha\beta}|=2^{N_l} \prod_i x^{L_i}
\label{eq:overlap}
\end{equation}
where $N_l$ is the number of loops in the transition graph; the product is
over these loops, $L_i$ being the length of the $i$th loop; and
$x={1}/{\sqrt{2}}$.  Thus, the overlap between two arbitrary
dimer coverings will often be small, as the transition graph will contain many
long loops, though (for a finite system) never zero.  However, there is
also a notion of a maximal overlap which occurs between two dimer coverings
that differ by only one minimal length loop.

This latter observation is the basis of the \emph{overlap expansion}, which is
an approximation
scheme based on treating $x$ in Eq.~(\ref{eq:overlap}) as a small parameter.\cite{Rokhsar88}
For example, the
overlap matrix of a set of square lattice dimer coverings, to leading order in
this expansion, is:
\begin{equation}
\Omega_{\alpha\beta}^{\text{square}}\approx \delta_{\alpha\beta} - 2x^4
\square_{\alpha\beta} + \dots
\label{eq:sqover}
\end{equation}
where $\square_{\alpha\beta}$ equals 0 unless the dimer coverings
$|d_\alpha\rangle$ and $|d_\beta\rangle$ differ by exactly one minimal length
loop, which on the square lattice has length four.  In this case,
$\square_{\alpha\beta}=\pm 1$, where the sign depends on the sign convention
we take for the singlets (i.\ e.\  our choice in whether a singlet between
spins $i$ and $j$ is written as $\frac{1}{\sqrt{2}}(i_\uparrow j_\downarrow -
i_\downarrow j_\uparrow)$ or $-\frac{1}{\sqrt{2}}(i_\uparrow j_\downarrow -
i_\downarrow j_\uparrow)$).  It can be shown that for any lattice, we can
choose the sign convention so that the overlap of two dimer coverings
differing by exactly one minimal loop always comes with a negative
sign.\cite{Raman05}  For the square lattice, this means that the
entries of the matrix $\square$ are always 0 or 1.  We will assume this sign
convention in this paper.

The overlap expansion can be applied to a general operator $\mathcal{O}$.  For
example, on a square lattice:
\begin{equation}
\mathcal{O}_{\alpha\beta}\equiv\langle d_\alpha |\mathcal{O}|d_\beta\rangle
\approx A_\alpha \delta_{\alpha\beta} - 2x^4 B_{\alpha\beta}
\square_{\alpha\beta}+\dots \label{eq:sqoper}
\end{equation}
where $A_\alpha = \mathcal{O}_{\alpha\alpha}$ and $B_{\alpha\beta}$ are
constants.  Eq. (\ref{eq:sqoper}) is especially convenient when $A_\alpha$ and
$B_{\alpha\beta}$ are independent of $\alpha,\beta$.

We will use the overlap expansion in the calculations that follow, but we
remind the reader that since $x$ is actually ${1}/{\sqrt{2}}$, the
approximation is a poorly controlled one.  Nonetheless,
the expansion has proven to be a useful guiding principle in the construction
of effective models of valence bond dominated phases.\cite{Rokhsar88, Elser93, Misguich02, Misguich03} We expect that for lattices where the minimal loop is large, such as the
length six loop of the kagom\'{e}, the approximation should work fairly well at leading order.

\subsection{Perturbation theory for susceptibility}
\label{sec:perturbation}

We now discuss our assumptions about the unperturbed
system (i.e. the Hamiltonian in the absence of DM interactions).
Then, armed with the technical background of the previous section, we shall derive
a perturbation theory for the susceptibility in the presence of a small DM interaction.

\begin{figure}[ht]
\includegraphics[width=\columnwidth]{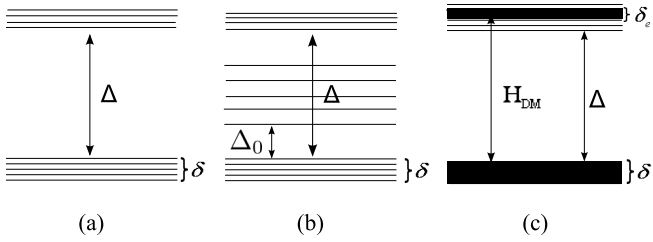}
\caption{The generality of our first assumption allows us to consider different possibilities for the spectrum of the unperturbed Hamiltonian. (a) The low energy sector is comprised of a narrow band of states which are superpositions of dimer coverings, separated by a spin gap $\Delta$ from the excited states.  (b) The same spectrum as in (a), but allowing a number of low-lying singlet states to exist in the gap. (c) Our second assumption:  We assume $H_\text{DM}$ connects the narrow band of low energy $S=0$ states with a narrow band of higher energy $S=1$ states.}
\label{fig:spec}
\end{figure}
\subsubsection{Assumptions}

Our \emph{first assumption} is that the spectrum of the unperturbed Hamiltonian has a low energy
structure resembling Fig.~\ref{fig:spec}a.  The figure depicts a set of low lying states
$\{ |n\rangle \}$, each of which can be written as a superposition of dimer coverings:
\begin{equation}
|n\rangle=\sum_\alpha a_{n\alpha} |d_\alpha\rangle. \label{eq:states}
\end{equation}
These states form a narrow band of width $\delta$ and are separated from magnetic
excitations by a spin gap $\Delta\gg\delta$.  While having a spin gap is crucial to the
analysis, our formalism can be adapted to a situation where higher energy singlet
states occur in the gapped region, as depicted in Fig.~\ref{fig:spec}b.  A well-known class
of toy models having (or widely believed to have) this structure are those of
the Klein--AKLT type.\cite{Klein82, Affleck87, Raman05, Nussinov07a}  These are models
for which dimer coverings, or a subset of them, are zero energy ground states and the
collection of coverings $\{ |d_\alpha\rangle \}$ forms a spin-gapped degenerate
ground state manifold.  In fact, we expect Fig.~\ref{fig:spec}ab to be a reasonably
accurate caricature of a spin-gapped system even when there is no obvious reason for
restricting our attention to nearest-neighbor valence bonds.  We will discuss the
suitability of this assumption (and the others) for the KHAF in section \ref{sec:kagome} but
we refer the interested reader to the summary in section IIF of Ref.~[\onlinecite{Misguich03}].

Our \emph{second assumption} is that $H_\text{DM}$ connects our narrow band of low energy
singlet states $\{ |n\rangle \}$ with a band of $S=1$ excited states $\{ |e_\gamma \rangle\}$, whose bandwidth $\delta_e$ is also small compared to $\Delta$, as depicted in Fig.~\ref{fig:spec}c.
While it is well-known that $H_\text{DM}$ will mix the singlet and triplet sectors, we are assuming
that the excited states $\{ |e_\gamma \rangle\}$, for which the matrix elements $\langle d_\alpha|H_\text{DM}|e_\gamma \rangle\neq 0$, are specifically $S=1$ eigenstates.  The simplest way to meet this requirement is via our \emph{third} assumption: that the unperturbed Hamiltonian conserves total spin, which both the Klein and Heisenberg models do.  We also assume these states form a narrow band.  One way this condition may arise is if the magnetic excitations can be viewed as local disturbances of the low energy states.
The simplest example of such a disturbance would be to break a single dimer by exciting the singlet to a triplet.

The precise way in which each of these assumptions are used will be made explicit during the derivation.

\subsubsection{Derivation}

Because the DM Hamiltonian (\ref{eq:DM}) does not commute with the total spin operator, $\mathbf{S}=\sum_i \mathbf{S}_i$, the sum being over sites in the lattice, deriving an expression for the susceptibility requires some care.  Consider the Hamiltonian:
\begin{equation}
H=H_0 + H_\text{DM} - g\mu_B \mathbf{h}\cdot\mathbf{S}
\label{eq:totalH}
\end{equation}
where $H_0$ is the unperturbed Hamiltonian discussed in the previous section, $\mathbf{h}$ is the magnetic field, $g$ the g-factor and $\mu_B$ the Bohr magneton.  If $\{ E_\alpha, |\alpha\rangle \}$ are
the eigenvalues and eigenstates of $H$, then the partition function is:
\begin{equation}
Z[T,\mathbf{h}]=\sum_\alpha e^{-E_\alpha(\mathbf{h})/k_B T}
\label{eq:Z}
\end{equation}
where $k_B$ is the Boltzmann constant and $T$ the temperature.  The magnetization $\mathbf{M}$ may be computed in the usual way:
\begin{eqnarray}
M^{\mu} &\equiv& -\frac{\partial F}{\partial h_\mu} = -\frac{\partial}{\partial h_\mu} \left ( -k_B T \ln Z \right ) =\frac{k_B T}{Z} \frac{\partial Z}{\partial h_\mu} \nonumber \\ &=& -\frac{1}{Z} \sum_\alpha \frac{\partial E_\alpha}{\partial h_\mu} e^{-E_\alpha/k_B T} \label{eq:mag}
\end{eqnarray}
and, similarly, the susceptibility:
\begin{eqnarray}
\chi^{\mu\nu} &\equiv& \frac{\partial M^\mu}{\partial h_\nu} = k_B T \left (\frac{1}{Z}\frac{\partial^2 Z}{\partial h_\mu \partial h_\nu} - \frac{1}{Z^2}\frac{\partial Z}{\partial h_\mu}\frac{\partial Z}{\partial h_\nu} \right ) \nonumber\\  &=& -\frac{1}{Z}\sum_\alpha \left (\frac{\partial^2 E_\alpha}{\partial h_\mu \partial h_\nu} \right  )e^{-E_\alpha/k_B T} \nonumber \\ &+& \frac{1}{k_B T}\Biggl [\frac{1}{Z}\sum _\alpha \left (\frac{\partial E_\alpha}{\partial h_\mu}\frac{\partial E_\alpha}{\partial h_\nu}\right ) e^{-E_\alpha/k_B T} \nonumber \\ &-& \frac{1}{Z^2}\sum _\alpha \frac{\partial E_\alpha}{\partial h_\mu} e^{-E_\alpha/k_B T} \sum _\beta \frac{\partial E_\beta}{\partial h_\nu} e^{-E_\beta/k_B T}\Biggr ] \nonumber \\ \label{eq:susc}
\end{eqnarray}

To compare with experiments, we need the powder susceptibility which is given by
$\chi_{\text{powder}}=\frac{1}{3}(\chi^{xx}+\chi^{yy}+\chi^{zz})$ or
\begin{multline}
\chi_\text{powder} = -\frac{1}{3Z}\sum_\alpha (\nabla^2_\mathbf{h} E_\alpha) e^{-E_\alpha/k_B T}
\\ +\frac{1}{3k_B T}\Bigl [\frac{1}{Z} \sum_\alpha (\nabla_\mathbf{h} E_\alpha)^2 e^{-E_\alpha/k_B T}
\\ -\frac{1}{Z^2}\Bigl (\sum_\alpha \nabla_\mathbf{h} E_\alpha e^{-E_\alpha/k_B T}\Bigr )^2\Bigr ]
\label{eq:suscpowder}
\end{multline}
where $\nabla_\mathbf{h} \equiv (\frac{\partial}{\partial h_x},\frac{\partial}{\partial h_y},\frac{\partial}{\partial h_z})$.

If the total spin $\mathbf{S}$ commutes with $H$, the states $\{ |\alpha\rangle \}$ may be chosen
as simultaneous eigenstates of the two operators and the eigenvalues will depend linearly on the magnetic field, i.e. $E_\alpha \sim \mathbf{h}\cdot\mathbf{S}_\alpha = \mathbf{h}\cdot\langle\alpha|\mathbf{S}|\alpha\rangle$.  In this case, the second derivative term on the RHS of Eq.~(\ref{eq:susc}) will vanish and the remaining terms will give $T\chi^{\mu\nu}\sim [\langle S^\mu S^\nu \rangle - \langle S^\mu\rangle\langle S^\nu\rangle]$, which is a familiar version of the fluctuation-dissipation theorem.
However, if $\mathbf{S}$ does not commute with $H$, the field dependence of the eigenvalues can
be more complicated.  For example, if the leading field dependence is quadratic in $\mathbf{h}$, then
in the zero field limit, the term in square brackets on the RHS of Eq.~(\ref{eq:susc}) will vanish and the
second derivative term will be all that remains.\footnote{Of course, the fluctuation-dissipation theorem will still be true.  However, the relationship between the uniform susceptibility and spin-spin correlation
will now involve more than just the equal time correlator.\cite{Auerbach98}}

Our first assumption was that the unperturbed Hamiltonian $H_0$ has a spectrum of the form shown in
Fig.~\ref{fig:spec}.  Because we also assumed that $H_0$ commutes with $\mathbf{S}$, the eigenstates of $H_0$ can be chosen to describe the system in the presence of a magnetic field.  In addition, if $g\mu_B h$ is sufficiently small compared to $\Delta$, the low energy spectrum of the Hamiltonian $H_1=H_0-g\mu_B \mathbf{h}\cdot\mathbf{S}$ will still resemble Fig.~\ref{fig:spec} and the lowest energy eigenstates will still be the collection of $S=0$ states $\{ |n\rangle \}$ of Eq.~(\ref{eq:states}).  However, the $S\neq 0$ bands will split into separate bands indexed by the spin component $S^h$ along the field direction $\hat{\mathbf{h}}$. In particular, the triplet band $\{ |e_\gamma\rangle \}$ mentioned earlier will split into three bands $\{ |e^{(S^h)}_\gamma\rangle \}$, where $S^h=-1,0,1$.  The spin gaps of the three bands are given by $\Delta^{(S^h)}=\Delta - g\mu_B h S^h$.  From Eq.~(\ref{eq:suscpowder}), one may verify that at low temperatures, the zero field susceptibility will vary as $\chi_\text{powder}\sim T^{-1} e^{-\Delta/T}$ which decreases to zero as $T\rightarrow 0$ as expected for a spin gapped system.

Having formally included the magnetic field exactly (at least with regard to the low energy spectrum), we now consider the effect of a small DM interaction, Eq.~(\ref{eq:DM}), on this picture.  If $D=|\mathbf{D}|$ is sufficiently small (a sufficient, though not necessary, condition for ``small" is if $D$ is small compared to the smallest of the three spin gaps $\Delta^{(S^h)}$), the spectrum of the perturbed Hamiltonian $H$ will still resemble Fig.~\ref{fig:spec} in the sense of a set of low lying states separated by a gap, though, since $H$ no longer conserves spin, the gap is no longer a ``spin" gap.  To determine the effect on the low temperature susceptibility, we need to examine how the low energy eigenvalues and eigenstates get modified through mixing with $S\neq 0$ components.

\begin{figure}[ht]
\includegraphics[width=0.8\columnwidth]{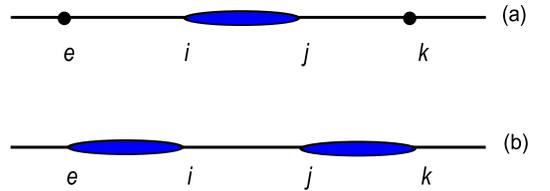}
\caption{Two different situations for a term in ${H}_{\text{DM}}$ to operate on: (a) a
dimer; (b) an empty bond.}
\label{fig:twobnds}
\end{figure}

To first order in perturbation theory, the states $\{ |n\rangle \}$ become:
\begin{equation}
|n'\rangle=|n\rangle +  \sum_{k\neq n}\frac{\langle
k|{H}_\text{DM}|n\rangle}{E_n-E_k}|k\rangle \label{eq:npert}
\end{equation}
where the sum is over all eigenstates of $H_1=H_0-g\mu\mathbf{h}\cdot\mathbf{S}$
except $|n\rangle$.  The sum will get restricted by our assumption that $H_\text{DM}$
connects the low energy sector $\{ |n\rangle \}$ with a narrow band of $S=1$ excited
states $\{ |e_\gamma\rangle \}$, split by the field into three separate narrow bands.
Moreoever, because the widths of these excited bands and of the low energy sector were
assumed to be small compared to the gap, we may approximate the
denominators by $E_k - E_n\approx \Delta^{(S^h)}$.  In this case:
\begin{eqnarray}
|n'\rangle &=&\sum_\alpha a_{n\alpha}\Biggl(|d_\alpha\rangle 
\nonumber\\ 
& &-
\sum_{S^h}\frac{1}{\Delta^{(S^h)}}\sum_{\gamma}{\langle
e^{(S^h)}_\gamma|{H}_\text{DM}|d_\alpha\rangle}|e^{(S^h)}_\gamma\rangle\Biggr)
\nonumber\\ 
&\equiv&
\sum_\alpha a_{n\alpha} |d_\alpha' \rangle.
\end{eqnarray}
Therefore, in this approximation, the consequence of having ${H}_{\text{DM}}$
present can be visualized in terms of its
effect on the individual dimer coverings:
\begin{equation}
|d_\alpha'\rangle=|d_\alpha\rangle - \sum_{S^h}\frac{1}{\Delta^{(S^h)}}\sum_{\gamma}{\langle
e^{(S^h)}_\gamma|{H}_\text{DM}|d_\alpha\rangle}|e^{(S^h)}_\gamma\rangle
\label{eq:dimerpert}
\end{equation}
To proceed, we now examine the effect of the operator ${H}_{\text{DM}}$ on one
of the dimer states.  As the operator is a sum over pairwise interactions,
${H}^{\langle ij \rangle}_{\text{DM}}$, there are
two cases to consider:  the link $\langle ij\rangle$ can either be (a)
occupied by a dimer or (b) an empty bond (see Fig.~\ref{fig:twobnds}).

We use the following notation to indicate the spin state of a pair of spins on
sites $i$ and $j$:
\begin{subequations}
\begin{equation}
(ij) \equiv \frac{1}{\sqrt{2}} \left(i_{\uparrow,n} j_{\downarrow,n} - i_{\downarrow,n}
j_{\uparrow,n}\right)
\label{eq:singlet}
\end{equation}
\begin{equation}
{[ij]^{n}_{0}} \equiv \frac{1}{\sqrt{2}} \left(i_{\uparrow,n} j_{\downarrow,n} + i_{\downarrow,n}
j_{\uparrow,n}\right)\\
\end{equation}
\begin{equation}
{[ij]^{n}_{1}} \equiv i_{\uparrow,n} j_{\uparrow,n}\\
\end{equation}
\begin{equation}
{[ij]^{n}_{-1}} \equiv i_{\downarrow,n} j_{\downarrow,n}
\end{equation}
\end{subequations}
-- the singlet and three triplets associated with the quantization axis $\hat{\mathbf{n}}$
where $i_{\uparrow(\downarrow),n}$ denotes the $S^n_i=\frac{1}{2} (-\frac{1}{2})$
eigenstate of the operator $\hat{S^n_i}$ and the subscripts $-1,0,1$ indicate the
value of $S^n_i+S^n_j$ for the pair $\langle ij\rangle$.  The notation
for a singlet $(ij)$ does not have a superscript because the singlet state is independent
of the choice for $\hat{\mathbf{n}}$.

The action of $H_\text{DM}$ on a dimer covering is most easily calculated with respect
to $\hat{\mathbf{z}}$ quantization axis.  Using the above notation, the action of the
term ${H}^{\langle ij\rangle}_{\text{DM}}$ in case (a), when spins $i$ and $j$
are in a singlet is:
\begin{multline}
{H}^{\langle ij\rangle}_{\text{DM}}(ij)=
\frac{D^z}{2i}[ij]^z_{0}-\frac{D^{-}}{2\sqrt{2}i}[ij]^z_{1}+\frac{D^{+}}{2\sqrt{2}i}[
ij]^z_{-1};
\label{eq:occ}
\end{multline}
while for case (b), when $\langle ij\rangle$ is an empty bond, the result is:
\begin{eqnarray}
\lefteqn{{H}^{\langle ij \rangle}_{\text{DM}}((ei)
(jk))=\frac{D^z}{4i}\left\{[ei]^z_{1}[jk]^z_{-1}-[ei]^z_{-1}[jk]^z_{1}
\right\}}
\nonumber\\
&& {}-\frac{D^{-}}{4\sqrt{2}i}\left\{[ei]^z_{1}[jk]^z_{0}-[ei]^z_{0}[jk]^z_{1}\right\}\nonumber\\
&& {}-\frac{D^{+}}{4\sqrt{2}i}\left\{[ei]^z_{-1}[jk]^z_{0}-[ei]^z_{0}[jk]^z_{-1}
\right\}
\label{eq:empty1} \\
&& =
-\frac{D^z}{4i}\left\{[ek]^z_0(ij)+(ek)[ij]^z_0\right\}\nonumber\\
&& {}+\frac{D^{-}}{4\sqrt{2}i}\left\{[ek]^z_1(ij)+(ek)[ij]^z_1\right\}\nonumber\\
&& {}-\frac{D^{+}}{4\sqrt{2}i}\left\{[ek]^z_{-1}(ij)+(ek)[ij]^z_{-1}\right\}.
\label{eq:empty2}
\end{eqnarray}
To avoid clutter, we have dropped the subscript $ij$ on the DM coefficients and
$D^{\pm}\equiv D^{x}\pm iD^{y}$.  Eqs.~(\ref{eq:occ}) and
(\ref{eq:empty1}) show that the effect of operator ${H}^{\langle ij\rangle}_{\text{DM}}$ on
a dimer covering is to promote the dimer(s) emanating from sites $i$
and $j$ from singlet(s) to triplet(s).  As indicated explicitly in Eqs.~(\ref{eq:occ}) and
(\ref{eq:empty2}), the state ${H}^{\langle ij\rangle}_{\text{DM}}|d_\alpha\rangle$, and hence
the state ${H}_{\text{DM}}|d_\alpha\rangle$, is an eigenstate of total spin with $S=1$.
An immediate consequence of this is that $\langle n|H_\text{DM}|n\rangle$,
which is a sum of matrix elements of the form
$\langle d_\alpha|H_\text{DM}|d_\beta \rangle$, is exactly zero.  Therefore,
the energy of state $|n\rangle$ will modified by $H_\text{DM}$ only at second
order in perturbation theory.

The second order correction to the energy of state $|n\rangle$ is given by:
\begin{multline}
E^{(2)}_n = \sum_{k\neq n} \frac{|\langle k|H_\text{DM}|n\rangle|^2}{E_n-E_k}
\\ \approx -\sum_{S^h}\frac{1}{\Delta^{(S^h)}}\sum_{\alpha,\beta,\gamma} a^*_{n\alpha} a_{n\beta}\langle d_\alpha|H_\text{DM}|e^{(S^h)}_\gamma\rangle\langle e^{(S^h)}_\gamma|H_\text{DM}|d_\beta\rangle\\
\label{eq:energypert2}
\end{multline}
The operators $\mathcal{P}^{(S^h)}\equiv \sum_\gamma |e^{(S^h)}\rangle\langle e^{(S^h)}|$  are
projection operators that respectively select for the $S^h=-1$, 0, and 1 components of $H_\text{DM}|d_\beta\rangle$.  Therefore, it is useful to rewrite Eqs.~(\ref{eq:occ})--(\ref{eq:empty2}) with respect to the two--spin states of the $\hat{\mathbf{h}}$ quantization axis.  If:
\begin{equation}
\hat{\mathbf{h}}=\sin\theta\cos\phi\hat{\mathbf{x}}+ \sin\theta\sin\phi\hat{\mathbf{y}} + \cos\theta\hat{\mathbf{z}}
\end{equation}
then Eq.~(\ref{eq:occ}) becomes:
\begin{multline}
{H}^{\langle ij\rangle}_{\text{DM}}(ij)=
\frac{D^h}{2i}[ij]^h_{0}-\frac{D^{\perp -}}{2\sqrt{2}i}[ij]^h_{1}+\frac{D^{\perp +}}{2\sqrt{2}i}[
ij]^h_{-1};
\label{eq:occ2}
\end{multline}
and similarly Eqs.~(\ref{eq:empty1}) and (\ref{eq:empty2}) become:
\begin{eqnarray}
\lefteqn{{H}^{\langle ij \rangle}_{\text{DM}}((ei)
(jk))=\frac{D^h}{4i}\left\{[ei]^h_{1}[jk]^h_{-1}-[ei]^h_{-1}[jk]^h_{1}
\right\}}
\nonumber\\
&& {}-\frac{D^{\perp -}}{4\sqrt{2}i}\left\{[ei]^h_{1}[jk]^h_{0}-[ei]^h_{0}[jk]^h_{1}\right\}\nonumber\\
&& {}-\frac{D^{\perp +}}{4\sqrt{2}i}\left\{[ei]^h_{-1}[jk]^h_{0}-[ei]^h_{0}[jk]^h_{-1}
\right\}
\label{eq:empty3} \\
&& =
-\frac{D^h}{4i}\left\{[ek]^h_0(ij)+(ek)[ij]^h_0\right\}\nonumber\\
&& {}+\frac{D^{\perp -}}{4\sqrt{2}i}\left\{[ek]^h_1(ij)+(ek)[ij]^h_1\right\}\nonumber\\
&& {}-\frac{D^{\perp +}}{4\sqrt{2}i}\left\{[ek]^h_{-1}(ij)+(ek)[ij]^h_{-1}\right\}.
\label{eq:empty4}
\end{eqnarray}
where
\begin{multline}
D^h=\mathbf{D}\cdot\hat{\mathbf{h}}=D^x\sin\theta\cos\phi+D^y\sin\theta\sin\phi+D^z\cos\theta
\label{eq:dh}
\end{multline}
\begin{multline}
D^{\perp \pm}=-D^z\sin\theta+D^x (\cos\theta\cos\phi\mp i\sin\phi)\\ \pm iD^y(\cos\phi\mp i\cos\theta\sin\phi)
\label{eq:dperp}
\end{multline}
The $\perp$ superscript refers to the direction perpendicular to the magnetic field and $D^{\perp +}$ and $D^{\perp -}$ are complex conjugates as the notation suggests.  From Eqs.~(\ref{eq:occ2}) and (\ref{eq:empty4}), we see that the amplitudes of the $S^h=0$ component of $H_\text{DM}|d_\beta\rangle$ are determined by the set of $\{ D^h_{ij} \}$ while the amplitudes of the $S^h=\pm 1$
components are determined by the collections $\{ D^{\perp\mp}_{ij} \}$.

The operator $H_\text{DM}\mathcal{P}^{(S^h)}H_\text{DM}$ of Eq.~(\ref{eq:energypert2}) is a sum of link terms such as $H_\text{DM}^{\langle ab\rangle}\mathcal{P}^{(S^h)}H_\text{DM}^{\langle cd\rangle}$.
The matrix element $\langle d_\alpha|H_\text{DM}^{\langle ab\rangle}\mathcal{P}^{(S^h=0)}H_\text{DM}^{\langle cd\rangle}|d_\beta\rangle$ will be proportional to:
\begin{equation}
D^h_{ab}D^h_{cd} = (\mathbf{D}_{ab}\cdot \hat{\mathbf{h}})(\mathbf{D}_{cd}\cdot \hat{\mathbf{h}})
\label{eq:dhprod}
\end{equation}
while $\langle d_\alpha|H_\text{DM}^{\langle ab\rangle}\mathcal{P}^{(S^h=\pm 1)}H_\text{DM}^{\langle cd\rangle}|d_\beta\rangle$ will be proportional to:
\begin{equation}
D^{\perp -}_{ab}D^{\perp +}_{cd} + D^{\perp +}_{ab}D^{\perp -}_{cd} = 2[\mathbf{D}_{ab}\cdot\mathbf{D}_{cd} - (\mathbf{D}_{ab}\cdot \hat{\mathbf{h}})(\mathbf{D}_{cd}\cdot \hat{\mathbf{h}})]
\label{eq:dperpprod}
\end{equation}
The proportionality constant (which might be zero) will depend on the overlap of states $|d_\alpha\rangle$ and $|d_\beta\rangle$ and the particular links $\langle ab\rangle$ and $\langle cd\rangle$ involved.

In order to proceed, we approximate Eq.~(\ref{eq:energypert2}) by the leading term in
its overlap expansion.  This means replacing the matrix $\langle d_\alpha|H_\text{DM}\mathcal{P}^{(S^h)}H_\text{DM}|d_\beta\rangle$ by the diagonal term in
(the lattice appropriate generalization of) Eq.~(\ref{eq:sqoper}).
We expect this approximation to be accurate if the mutual overlaps between the
dimer coverings entering the superposition in Eq.~(\ref{eq:states}) are small or
if the lattice architecture involves large minimal length loops.  In this approximation, Eq.~(\ref{eq:energypert2}) becomes:
\begin{multline}
E^{(2)}_n \approx -\sum_{S^h}\frac{1}{\Delta^{(S^h)}}\sum_{\alpha} |a_{n\alpha}|^2 \\
\times
\sum_{\langle ab\rangle, \langle cd\rangle} \langle d_\alpha|H^{\langle ab\rangle}_\text{DM}\mathcal{P}^{(S^h)}\mathcal{P}^{(S^h)}H^{\langle cd\rangle}_\text{DM}|d_\alpha\rangle
\\ + O(x^{L_\text{min}})
\label{eq:energypert3}
\end{multline}
where $L_\text{min}$ is the length of the minimal loop which can appear in a transition graph for
the lattice in question and, to the same order, $\sum_\alpha |a_{n\alpha}|^2 \approx 1 + O(x^{L_\text{min}})$.  We have also used a basic property of projection operators:  $\mathcal{P}^2=\mathcal{P}$.

\begin{figure}[h!]
\includegraphics[width=0.25\columnwidth]{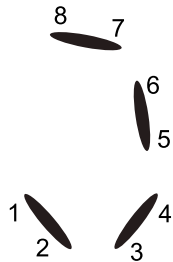}
\caption{Generic 2D arrangement considered for determining the distinct combinations of $H^2_\text{DM}$, as listed in Table~\ref{tab:gen}.  The specific Table~\ref{tab:gen} cases which arise
will depend on the connectivity of the lattice under consideration.}
\label{fig:genlat}
\end{figure}

For a dimer covering $|d_\alpha\rangle$, the set of vectors $\{ \mathcal{P}^{(S^h)}H_\text{DM}^{\langle ij\rangle}|d_\alpha\rangle \}_{\langle ij\rangle}$ can be determined by looking at Eqs.~(\ref{eq:occ2})--(\ref{eq:empty4}).  Eq.~(\ref{eq:energypert3}) involves the sum of all possible overlaps between pairs of vectors in this set.  These pairs can be classified into ten distinct types of combinations which are
listed in Table \ref{tab:gen} with reference to a dimer covering of the generic lattice depicted in Fig.~\ref{fig:genlat}.  The terms that actually arise will depend on the connectivity of the lattice under consideration.

The vector $\mathcal{P}^{(S^h)}H_\text{DM}^{\langle ab\rangle}|d_\alpha\rangle$ resembles $|d_\alpha\rangle$ except the dimer(s) emanating from sites $a$ and $b$ have been promoted to triplets.  Therefore, the diagonal matrix element $\langle d_\alpha|H_\text{DM}^{\langle cd\rangle}\mathcal{P}^{(S^h)}\mathcal{P}^{(S^h)}H_\text{DM}^{\langle ab\rangle}|d_\alpha\rangle=0$ except for when links $\langle ab\rangle$ and $\langle cd\rangle$ involve promoting the same dimer(s) to triplets.  This corresponds to cases 1, 5, 6, and 8 in Table \ref{tab:gen}.  The other cases in Table \ref{tab:gen} will contribute at higher orders in the overlap expansion, which involve off-diagonal matrix elements.

Using Eqs.~(\ref{eq:occ2})--(\ref{eq:empty3}), (\ref{eq:dhprod}), and (\ref{eq:dperpprod}),
Eq.~\ref{eq:energypert3}, to leading order in the overlap expansion, becomes:
\begin{multline}
E^{(2)}_n =  -\frac{1}{8\Delta}{\sum}^{(0)}_{\langle ab\rangle} \left(D^2_{ab} + \eta \left[D^2_{ab} - (\mathbf{D}_{ab}\cdot\hat{\mathbf{h}})^2\right]\right)\\
- \frac{1}{8\Delta}\sum_\alpha |a_{n\alpha}|^2 \Biggl\{
{\sum}^{(1)}_{(ab)\in |d_\alpha\rangle} \Bigl(D^2_{ab} \\
+ \eta \left[D^2_{ab} - (\mathbf{D}_{ab}\cdot\hat{\mathbf{h}})^2\right]\Bigr)\\
-2 {\sum}^{(2)}_{\triangle [a(bc)]\in |d_\alpha\rangle} \Bigl(\mathbf{D}_{ab}\cdot\mathbf{D}_{ac}
\\
+ \eta \left[\mathbf{D}_{ab}\cdot\mathbf{D}_{ac} - (\mathbf{D}_{ab}\cdot\hat{\mathbf{h}})(\mathbf{D}_{ac}\cdot\hat{\mathbf{h}})\right]\Bigr)\\
+2 {\sum}^{(3)}_{\square [(ab)(cd)]\in |d_\alpha\rangle} \Bigl(\mathbf{D}_{ad}\cdot\mathbf{D}_{bc}
\\
+ \eta \left[\mathbf{D}_{ad}\cdot\mathbf{D}_{bc} - (\mathbf{D}_{ad}\cdot\hat{\mathbf{h}})(\mathbf{D}_{bc}\cdot\hat{\mathbf{h}})\right]\Bigr)\Biggr\}\\ + O(x^{L_\text{min}})
\label{eq:energypert4}
\end{multline}
where $\eta = {\left[\left(\frac{\Delta}{g\mu_B h}\right)^2-1\right]^{-1}}$.
The first sum in Eq.~(\ref{eq:energypert4}), labelled with superscript (0), is over \emph{all} links
in the lattice and is a uniform shift that is the same for every state $|n\rangle$ in the low energy
sector.  The three sums, labelled with superscripts (1), (2), (3), are taken with reference to a
particular dimer covering $|d_\alpha\rangle$.  These terms will therefore contribute to
$E_n^{(2)}$ according to which dimer coverings enter the superposition in Eq.~(\ref{eq:states}).
The sum labelled (1) is over all links $\langle ab\rangle$ that contain a dimer.  The sum
labelled (2) is over all triangular plaquettes that contain a dimer, where $\triangle [a(bc)]$ means
that the dimer lives on link $\langle bc\rangle$.  The sum labelled (3) is over square plaquettes $\square [(ab)(cd)]$ with two dimers on links $\langle ab\rangle$ and $\langle cd\rangle$.  These last two sums will
only occur if the lattice in question contains triangular and square plaquettes respectively.

Eq.~(\ref{eq:energypert4}) gives the field dependence of the energy levels of the states in
the low energy manifold.  We expect  these low lying states to dominate the thermal averages of Eq.~(\ref{eq:susc}) when the temperature and magnetic field energy are small compared to
the gap $\Delta$ due to the Boltzmann factors.  If we therefore make the approximation of restricting
the sums of Eq.~\ref{eq:susc} to the low energy manifold, Eq.~\ref{eq:energypert4} tells us that
the leading field dependence (at low fields) is quadratic in $h$.  Therefore, the zero field susceptibility
will be determined by the second derivative terms of Eq.~\ref{eq:susc} which, as mentioned above,
vanish for systems where total spin is conserved.

To compare with experiments, we calculate the zero field powder susceptibility in the
$T\rightarrow 0$ limit:
\begin{multline}
\chi_\text{powder} = \frac{(g\mu_B)^2}{6\Delta^3}\Biggl\{ {\sum}^{(0)}_{\langle ab\rangle} D^2_{ab} \\ +  \frac{1}{Z}{\sum}'_n w_n \sum_\alpha |a_{n\alpha}|^2 \Bigl( {\sum}^{(1)}_{(ab) \in |d_\alpha\rangle} D^2_{ab} \\ - 2 {\sum}^{(2)}_{\triangle [a(bc)]\in |d_\alpha\rangle} \mathbf{D}_{ab}\cdot\mathbf{D}_{ac} \\
+ 2 {\sum}^{(3)}_{\square [(ab)(cd)]\in |d_\alpha\rangle} \mathbf{D}_{ad}\cdot\mathbf{D}_{bc} \Bigr) \Biggr\}
\label{eq:chipowder2}
\end{multline}
where $w_n = e^{-E_n/k_B T}$ is the Boltzmann factor of state $|n\rangle$ and the primed sum is
restricted to the low energy manifold.  The expression can be simplified if we assume the magnitude
of $\mathbf{D}$ is the same on every link and the products
$\mathbf{D}_{ab}\cdot\mathbf{D}_{ac} = D^2 \cos\theta_\triangle$ and
$\mathbf{D}_{ad}\cdot\mathbf{D}_{bc}=D^2\cos\theta_\square$ are the
same for every triangular and square plaquette respectively.  In this case, Eq.~\ref{eq:chipowder2} becomes:
\begin{multline}
\chi_\text{powder}= \frac{N(g\mu_B)^2 D^2}{6\Delta^3}
\left(\frac{z+1}{2} -2\cos\theta_\triangle \frac{\langle N_\triangle\rangle}{N}
\right.
\\
\left.
+ 2\cos\theta_\square \frac{\langle N_\square\rangle}{N}\right)
\label{eq:chipowder3}
\end{multline}
where $z$ is the coordination of the lattice, $N$ the number of sites, and
$\langle N_{\triangle,\square}\rangle \equiv \frac{1}{Z}{\sum}' w_n \sum_\alpha |a_{n\alpha}|^2 N_{(\triangle,\square),\alpha}$ where $N_{(\triangle,\square),\alpha}$ is the number of triangular
(square) plaquettes which contain one (two) dimer(s) in dimer covering $|d_\alpha\rangle$.

\begin{table}[h]
\begin{tabular}{p{0.5cm} p{2.5cm}  p{0.5cm} l}
\hline
\hline
1. & $H_\text{DM}^{12} \mathcal{P}^{(S^h)} H_\text{DM}^{12}$
   & 6. & $H_\text{DM}^{24}\mathcal{P}^{(S^h)} H_\text{DM}^{23}$\\
\hline
2. & $H_\text{DM}^{34} \mathcal{P}^{(S^h)} H_\text{DM}^{12}$
   & 7. & $H_\text{DM}^{25}\mathcal{P}^{(S^h)} H_\text{DM}^{23}$\\
\hline
3. & $H_\text{DM}^{23} \mathcal{P}^{(S^h)} H_\text{DM}^{12}$
   & 8. & $H_\text{DM}^{14}\mathcal{P}^{(S^h)} H_\text{DM}^{23}$\\
\hline
4. & $H_\text{DM}^{45}\mathcal{P}^{(S^h)}  H_\text{DM}^{12}$
   & 9. & $H_\text{DM}^{15}\mathcal{P}^{(S^h)} H_\text{DM}^{23}$\\
\hline
5. & $H_\text{DM}^{23}\mathcal{P}^{(S^h)}  H_\text{DM}^{23}$
   & 10. & $H_\text{DM}^{67}\mathcal{P}^{(S^h)} H_\text{DM}^{23}$\\
\hline
\hline
\end{tabular}
\caption{Calculating the second order (in DM) correction to
the energy, $E_n^{(2)}$, involves computing all possible overlaps
of pairs of vectors in the set $\{ \mathcal{P}H_\text{DM}^{\langle ij\rangle}
|d_\alpha\rangle \}_{\langle ij\rangle}$ for each dimer covering in the superposition that
defines the unperturbed state $|n\rangle$ (Eq.~(\ref{eq:states})).  There are
ten distinct cases to consider, indexed in this table by the link operators involved,
in reference to the generic dimer covering in Fig.~\ref{fig:genlat}.  The possibilities are
both links involving dimers, (1)--(2);  one link involving a dimer, (3)--(4);  and neither
link involving a dimer (5)--(10).  If both links involve a dimer, they can be either (1) the same
or (2) different dimers.  If one link involves a dimer, then the other link is an empty bond that
either (3) contains or (4) does not contain one of the spins of that dimer.  If neither link involves
a dimer, then the possibilities are (5) the two empty links are the same;  the two links
share exactly one spin while the other two spins either (6) form a dimer --- this is possible if
the lattice has triangular plaquettes or (7) do not form a dimer; or the two links do not share
a spin (8)--(10).  In this latter case, the two empty links may be connected by
(8) two dimers --- this is possible if the lattice has square plaquettes;  (9) exactly one
dimer;  or (10) no dimers.}
\label{tab:gen}
\end{table}

Eq.~(\ref{eq:chipowder3}) shows explicitly that the low temperature, zero field powder
susceptibility does not  vanish at low temperatures but instead approaches a constant that
depends on $D$.  From Eqs.~(\ref{eq:mag}) and (\ref{eq:energypert4}), we find that quadratic
field dependence also implies that the low temperature, zero field magnetization will be
zero.  Therefore, the picture of a short-ranged valence bond phase with DM interactions
is qualitatively consistent with the combination of a non-vanishing $T\rightarrow 0$ susceptibility and lack
of magnetic order observed in experiments on the herbertsmithite compound.\cite{Helton07,Ofer07,Mendels07,Imai08}
In the following sections, we discuss how this approach fares quantitatively in the context of
various models where our assumptions are known or widely believed to be satisfied.

\section{Application to the Shastry--Sutherland model of $\text{SrCu}_2(\text{BO}_3)_2$}
\label{sec:ss}

\begin{figure}[h]
\includegraphics[width=0.9\columnwidth]{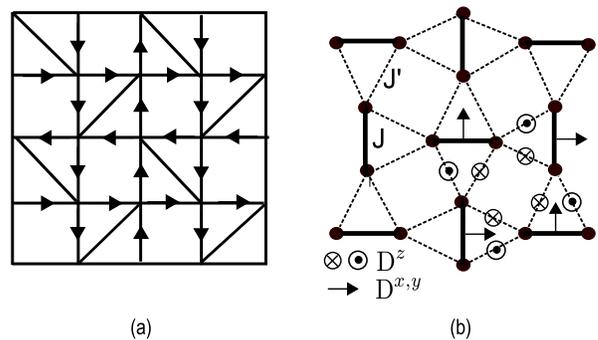}
\caption{(a) The Shastry--Sutherland (SS) lattice. In its ground state, the spins connected by the diagonal bonds form singlets. The arrows
on the links denote their orientations in terms of the sign convention for $H_\text{DM}$, i.e. which spin comes first in the cross product.  The orientation of the
diagonal bonds can be taken as going from the lower site to the upper site, though
this fact will not enter the calculation.  (b) The structure
formed by the copper ions in  $\text{SrCu}_{2}(\text{BO}_3)_2$. This is
topologically equivalent to the SS lattice.  Indicated in this figure are the directions
of the $\mathbf{D}$ vectors on the various links, where the orientation convention
of (a) has been assumed.}
\label{fig:shastry}
\end{figure}

We now consider a toy spin model defined on the lattice in Fig.~\ref{fig:shastry}a, which was first considered by Shastry and Sutherland (SS) in 1981.\cite{Shastry81}  This model has been revisited more recently\cite{Kageyama99, Miyahara99} in light of the spin-dimer compound $\text{SrCu}_2(\text{BO}_3)_2$, where the copper ions form a lattice (Fig.\ref{fig:shastry}b) which is topologically equivalent to the one considered by SS.  From our standpoint, this model satisfies the assumptions of section \ref{sec:model} almost exactly, which is why we discuss it as a first application of our formalism.

\subsection{Model}

In the SS model, the interaction between neighboring spins depends on whether the link
connecting them is horizontal, vertical, or diagonal.  The Hamiltonian is:
\begin{equation}
\label{eq:SSHam}
H_0 = J {\sum _{\langle ij\rangle}}^\text{diag}\mathbf{S}_i \cdot \mathbf{S}_j+J^{\prime} {\sum
_{\langle ij \rangle}}^{\text{horiz,vert}}\mathbf{S}_i \cdot \mathbf{S}_j.
\end{equation}
where $J$ and $J'$ are positive constants.  If $J=0$, the model reduces to the square
lattice Heisenberg antiferromagnet, which is widely believed to have a N\'{e}el ordered
ground state.  If $J'=0$, the ground state is a product state of singlets on the diagonal
bonds: $|\Psi\rangle=\prod^\text{diag}_{\langle ij \rangle}(ij)$.  SS showed that this product state, which we may call a valence bond solid,\footnote{We might view a valence bond solid as ``exotic'' if it occurs spontaneously due to quantum frustration. However, in the SS model, and experimentally in spin-dimer materials\cite{Kageyama99}, the valence bond solid phase occurs because some links have a stronger interaction than others.  The discussion in the introduction where we counted the valence bond solid as an example of an exotic non-magnetic phase, was in reference to the first scenario.} is an exact eigenstate of the full Hamiltonian (\ref{eq:SSHam}) and,
in fact, the ground state up to a critical value of ${J'}/{J}$, which they determined to be of order unity.  When $J'=0$, the basic magnetic excitations involve replacing singlets with triplets and the spin
gap is exactly equal to $J$.  In Ref.~[\onlinecite{Miyahara99}], it was shown that the excitations
remain nearly localized with a small dispersion when $J'\neq 0$, except for very close to the transition (in fact, it has been shown that even singlet excitations are gapped\cite{Cepas01} in this parameter range).  Therefore, in its valence bond phase, the SS model has a spectrum like Fig.~\ref{fig:spec}a, where the low energy sector now consists of just one state $|\Psi\rangle$.

We now augment Eq.~(\ref{eq:SSHam}) with a perturbatively small nearest-neighbor DM interaction.  For an ideal 2D lattice, we only need to consider the component $D^z$.  Moreover, for an ideal SS lattice, $\mathbf{D}=0$ on the diagonal bonds because the midpoints of those bonds are centers of inversion symmetry.\cite{Cepas01}  However, in the context of $\text{SrCu}_2(\text{BO}_3)_2$, it has been argued\cite{Miyahara04} that a slight buckling of the planes removes these symmetries and the leading effect is to induce $D^x$ and $D^y$ terms on the diagonal bonds.  Therefore, in anticipation of relating to the experiment, we consider the DM interactions shown in Fig.~\ref{fig:shastry}.

In adapting Eq.~(\ref{eq:chipowder2}) to this model, a few points should be noted.  First is the fact that  $H_\text{DM}$ couples the ground state to two different bands of magnetic excitations (each of which gets split into three bands on application of a magnetic field).  To see this, it is simplest to consider the $J'=0$ case first.  Referring to Table \ref{tab:gen}, we see that case 1 terms connect $|\Psi\rangle$ to the set of magnetic eigenstates where one of the dimers is now a triplet.  These states have an energy $\Delta_1=J$ above the ground state.  Cases 5 and 6 connect $|\Psi\rangle$ to eigenstates where two of the dimers are triplets (case 8 terms, while not forbidden by the lattice, do not occur in our considered ground state).  These states have energy $\Delta_2=2J$ above the ground state.  Since these two bands are orthogonal to one another, we can consider them separately in the calculation.  In particular, the argument which used the assumption of having a narrow band, can be easily generalized to the present case because both of these bands are narrow.
For instance, one may verify that the proper generalization of Eq.~(\ref{eq:dimerpert}) is:
\begin{multline}
|\Psi'\rangle=|\Psi\rangle -  \sum_{S^h}\frac{1}{\Delta_1^{(S^h)}}{\sum_{\gamma}}'{\langle
e_\gamma^{(S^h)}|{H}_\text{DM}|\Psi\rangle}|e_\gamma^{(S^h)}\rangle \\ -\sum_{S^h}
\frac{1}{\Delta_2^{(S^h)}}{\sum_{\gamma}}''{\langle
e_\gamma^{(S^h)}|{H}_\text{DM}|\Psi\rangle}|e_\gamma^{(S^h)}\rangle.
\label{eq:psipert}
\end{multline}
where the first sum is over the $S=1$, $E=\Delta_1=J$ band while the second is over the $S=1$,  $E=\Delta_2=2J$ band.  These arguments will still apply for the case where $J'\neq 0$ (when we are away from the transition point) though the bands will now acquire a small dispersion and the spin gaps will be renormalized\cite{Miyahara99, Cepas01} (and $\Delta_2$ will no longer be $2\Delta_1$).  We will account for this in our calculation by using the actual gap values $\Delta_{1,2}$ in Eq.~(\ref{eq:psipert}).

Similarly, the equation for the powder susceptibility will now consist of two separate parts, each of the form in Eq.~\ref{eq:chipowder2}, corresponding to the two bands of excited states.  For the term associated with the $\Delta_1$ band, $\sum^{(0)} D_{ab}^2 = \sum^{(1)} D_{ab}^2 = {N} \left[(D^x)^2+(D^y)^2\right]/4$ while the other two sums do not contribute.  For the term associated with the $\Delta_2$ band,
$\sum^{(0)} D_{ab}^2 = 4N_d(D^z)^2=2N(D^z)^2$ while the second sum will not contribute because
in our chosen state, the dimers are only on diagonal links.  One may verify that each triangle contributing
to $\sum_{(2)} \mathbf{D}_{ab}\cdot\mathbf{D}_{ac}$ contributes the same value, $\mathbf{D}_{ab}\cdot\mathbf{D}_{ac} = -(D^z)^2$ and the number of these triangles is $2N_d=N$.  Therefore, we arrive at our final result:
\begin{multline}
\chi_\text{powder}\approx \frac{N(g\mu_B)^2}{12}\left(\frac{(D^x)^2+(D^y)^2}{\Delta_1^3} + \frac{8(D^z)^2}{\Delta_2^3}\right)\\
\label{eq:chiss}
\end{multline}
A noteworthy point is that because the low energy sector of the unperturbed model contains only one state which happens to be a dimer covering, higher order terms in the overlap expansion will not enter.  Therefore, the only approximations in Eq.~(\ref{eq:chiss}) are those inherent in perturbation theory.

\subsection{Comparison with experiment}

The most striking feature of the temperature dependence of the zero field\footnote{The experiments we are referring to (Ref.~[\onlinecite{Kageyama99}]) were actually done in a magnetometer with a static field of $H=1.0$ T, which for spin 1/2 corresponds to an energy scale $g\mu_B H\sim 1.3$~K.  The temperature range of the experiment is from 1.7 to 400~K, so the results can be interpreted by a zero-field theory except for perhaps the very lowest temperature points.} susceptibility of $\text{SrCu}_2(\text{BO}_3)_2$ is a peak, which occurs around $T=15$~K, followed
by a rapid decrease to nearly zero as the temperature is further lowered.\cite{Kageyama99}  The
most natural interpretation of this rapid decrease is a reduction of entropy as the physics becomes
increasingly dominated by a non-magnetic ground state.  ESR measurements\cite{Nojiri99}
revealed two sets of triplet excitations with spin gaps $\Delta_1=35$~K and $\Delta_2=55$~K respectively.\footnote{Measurements of the NMR relaxation rate\cite{Kageyama99} determined the gap to
the lowest magnetic state to be $\Delta_1 = 30$~K, consistent with the value of $\Delta_1=35$~K given in Ref.~[\onlinecite{Nojiri99}].  Fitting the susceptibility data gave a slightly lower value of $\Delta = 19$~K but these and other authors\cite{Miyahara99} have interpreted the $\approx$30~K values as the ``spin gap".  Therefore, we use this value for our estimate.}

The SS model has had some success as a theoretical description of both the ground and excited states of $\text{SrCu}_2(\text{BO}_3)_2$ in zero magnetic field.  The analysis of Ref.~[\onlinecite{Miyahara99}] determined that the temperature dependence of the susceptibility was well modelled by Eq.~(\ref{eq:SSHam}) with a value of ${J'}/{J}\approx 0.68$, which is in the valence bond phase but somewhat close to the transition point (though still far enough away to have appreciable spin and singlet gaps and fairly localized triplet excitations\cite{Miyahara99, Cepas01}).
In addition, it was shown\cite{Cepas01, Miyahara04, Miyahara05} that by including DM interactions, the model could also explain a number of features that occurred in the presence of a magnetic field, such as the appearance of uniform and staggered magnetizations\cite{Kageyama99,Kodama99} in fields small compared to the spin gap scale.\footnote{The spin gap scale is $\sim {\Delta}/{g\mu_B} \sim 20$ T while the onset of a uniform magnetization is seen at around 18 T.  We point
out that these interesting field dependencies occur at fields much higher than the 1.0 T field of the magnetometer used in Ref.~[\onlinecite{Kageyama99}], the results of which we interpret with our zero field calculation.}

Eq.~(\ref{eq:chiss}) predicts the effect of DM in the case of zero magnetic field: as $T\rightarrow 0$,
the susceptibility will not decay to zero but instead approach a constant.  To compare with experiment, we use $g\approx 2$, the measured values for $\Delta_1\approx 35$~K and $\Delta_2\approx 55$~K given in Ref.~[\onlinecite{Nojiri99}], an estimate of $D^z\approx 2$~K taken from ESR measurements\cite{Cepas01}, and the suggestion of Ref.~[\onlinecite{Miyahara04}]  that $D^{x,y}\sim D^z$.  The molar susceptibility is obtained by multiplying Eq.~(\ref{eq:chiss}) by the factor ${N_A}/{N}$, where $N_A$ is Avogadro's number.  With these numbers, we obtain the estimate: $\chi_\text{powder}(T=0)\approx 5\times 10^{-5}$ emu/mol Cu.  The measured susceptibility in Fig. 2 of Ref.~[\onlinecite{Kageyama99}] does not go to zero and, in fact, begins to show a rise at a temperature after reaching a minimum value of around $5\times 10^{-4}$ emu/mol Cu at roughly 4~K, which the authors attributed to a small ($< 1\%$) concentration of impurities.

Therefore, more experiments on substantially cleaner samples are
needed to test Eq.~(\ref{eq:chiss}).  We reemphasize that the assumptions of section \ref{sec:model} are expected to hold well for the SS model so to the extent that Eq.~(\ref{eq:SSHam}) with DM included is a good model of $\text{SrCu}_2(\text{BO}_3)_2$, we expect our order of magnitude estimate to be reliable.

\section{Application to the generalized Klein model on the checkerboard lattice}
\label{sec:check}

\begin{figure}[ht]
\includegraphics[width=0.85\columnwidth]{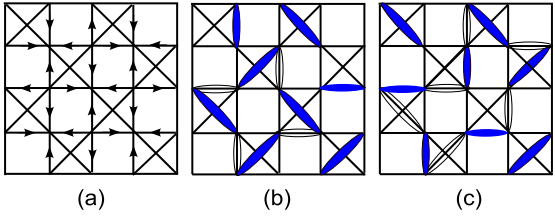}
\caption{(a) The checkerboard lattice with arrows specifying the orientations
of links with DM interactions.  On horizontal and vertical links, we take
$\mathbf{D}=D^z\hat{z}$ (i.e. the sign is always positive) while on diagonal links $\mathbf{D}=0$.
(b) The minimal loop on this lattice has length 8.
(c) The next minimal loop has length 12.}
\label{fig:check}
\end{figure}

We next consider a toy model defined on the checkerboard lattice shown in Fig.~\ref{fig:check}a.
The sites of this lattice are the same as a square lattice but the connectivity is different:
on alternate square plaquettes, opposite corners are also connected as nearest-neighbors by diagonal links (note: the crossing point of these diagonal links is \textbf{not} an additional site).
The lattice can be viewed as a 2D projection of the 3D pyrochlore lattice.

The checkerboard lattice does not currently have a realization in an actual material.  While there are a number of materials\cite{Bramwell01} which form pyrochlore structures, we are not aware of a system where the low energy physics is believed to be captured by short-range valence bonds.  However, there is a theoretical motivation for considering this system. The model we are about to describe is among the simplest examples of a model for which (a) the Hamiltonian and exact ground state(s) are known and well-characterized, (b) the assumptions of section \ref{sec:model} are expected to either hold or be a decent approximation, and (c) the overlap expansion is expected to converge fairly well, while at the same time the model (d) captures all of the complexities of the most general scenario.  In this sense, we may consider this system as a ``laboratory" in which to explore the limitations of our formalism.

The model we consider is a two-dimensional version\cite{Nussinov07a} of the generalized Klein model, introduced in Ref.~[\onlinecite{Raman05}] (note: a model of a similar type was also considered in Ref.~[\onlinecite{Batista04}]).  The Hamiltonian for this model may be written as:
\begin{equation}
H=\sum_p \hat{h}_p \label{eq:KKHam}
\end{equation}
where the sum is over \emph{crisscrossed} square plaquettes and:
\begin{equation}
\hat{h}_p = (S_p^2)(S_p^2-2). \label{eq:hp}
\end{equation}
where $\mathbf{S}_p\equiv\mathbf{S}_1+\mathbf{S}_2+\mathbf{S}_3+\mathbf{S}_4$ is the total spin of the four members of plaquette $p$.  $\hat{h}_p$ is an operator that projects the wave function onto its component where plaquette $p$ has the maximal spin of 2, the two factors in Eq.~(\ref{eq:hp}) annihilating the spin 0 and 1 components respectively.  Since Eq.~(\ref{eq:KKHam}) is a sum of projection operators, its eigenvalues must be non-negative and any state with zero eigenvalue is a ground state.

A wave function in which crisscrossed plaquette $p$ has a dimer on one of its links will be annihilated by $\hat{h}_p$ since the total spin of $p$ can then be at most 1.  Therefore, a wave function having a dimer on \emph{every} crisscrossed plaquette will be a zero energy ground state of Eq.~(\ref{eq:KKHam}).  A counting argument given in Refs.~[\onlinecite{Raman05}] and [\onlinecite{Nussinov07a}] shows that the ground state manifold of Eq.~(\ref{eq:KKHam}) consists \emph{only} of dimer coverings of the lattice where every crisscrossed plaquette has \emph{exactly} one dimer and superpositions of such coverings.  Note that there are dimer coverings of the checkerboard lattice that are not included in this set.

In terms of Fig.~\ref{fig:spec}a, the low energy sector of Eq.~(\ref{eq:KKHam}) is a degenerate manifold.  One may consider perturbing Eq.~(\ref{eq:KKHam}) to arrive at a model with a smaller ground state space.  It has not been proven, but strongly suspected, that the Hamiltonian~(\ref{eq:KKHam}) has a spin gap; demonstrations of a spin gap have been made, to varying degrees of rigor, in a number of related models.\cite{Majumdar69, Affleck87, Raman05, Caspers84}  Therefore, there are strong reasons to believe that Eq.~(\ref{eq:KKHam}) satisfies the assumptions required by our formalism.

Moreover, the loop structure of the ground state space is such that there are strong reasons to expect the overlap expansion to converge well in a wide number of situations.\footnote{One situation where the expansion might not work so well is if the wave function were a liquid state of the Rokhsar--Kivelson type.\cite{Rokhsar88}.  In this case, the error made in keeping only leading terms might be quite large because of the cumulative effect of many small terms.}  Because the minimal loop has length 8 (Fig.~\ref{fig:check}b), keeping only the diagonal term of the overlap expansion will often be a good approximation.  Also, the next minimal loop has length 12 (Fig.~\ref{fig:check}c), instead of 10, so keeping the leading off-diagonal term in the expansion will result in a relative error of $\sim x^{(12-8)}\sim x^4$, while in many other lattices, including the square, this error will be $\sim x^2$.

The simplest way of introducing a nearest-neighbor DM coupling is to consider an ideal 2D lattice, in which case we only have $D^z$ on the horizontal and vertical links ($\mathbf{D}=0$ on the diagonal links due
to inversion symmetry).  Because we do not have a specific material in mind, we take for simplicity $D^z$ to be the same magnitude and sign on all of the links, as per the link orientation convention in Fig.~\ref{fig:check}a.  In adapting Eq.~(\ref{eq:chipowder2}) to the present case, we note that all four cases contributing to that general expression will occur here. The first sum will give $\sum^{(0)} D_{ab}^2 = 2N (D^z)^2$, where $2N$ is total number of non-diagonal links.  Similarly, the second sum will give $\sum^{(1)} D_{ab}^2 = (D^z)^2 (\frac{N}{2} - N_\text{diag})$ where $N_\text{diag}$ is the number of diagonal dimers in the coverings under consideration.  There will be two triangular terms for each diagonal dimer, and for each of these terms $\mathbf{D}_{ab}\cdot\mathbf{D}_{ac} = (D^z)^2$.  Similarly,
each square plaquette having two dimers will contribute $-(D^z)^2$.  Therefore, the analog of Eq.~(\ref{eq:chipowder3}) for the present case is:
\begin{multline}
\chi_\text{powder}\approx \frac{5N(g\mu_B)^2(D^z)^2}{12 \Delta^3} \left [1-\frac{2\langle N_\text{diag} \rangle}{N}-\frac{4\langle N_\square\rangle}{5N} \right]\\ + O(x^8)
\label{eq:chiKK}
\end{multline}
where, $\langle\cdots\rangle$ denotes thermal average as per the discussion after Eq.~(\ref{eq:chipowder2}).

It would be useful to study Eq.~(\ref{eq:KKHam}) numerically to confirm the existence of a
spin gap and to examine its low-lying $S=1$ states, in light of the assumptions of our formalism.
A comparison of an ``exact" calculation of $\chi_\text{powder}$ for (possibly a perturbed version of) Eq.~(\ref{eq:KKHam}) with the analytical expression Eq.~(\ref{eq:chiKK}) would provide a useful validation of our approach and is natural topic for further study.

\section{Application to the Kagom\'{e} Lattice and $\text{ZnCu}_3\text{(OH)}_6\text{Cl}_2$}
\label{sec:kagome}

In the previous sections, we have applied our formalism to systems with
well-characterized Hamiltonians and ground states.  In the present section, we
finally return to the original motivation for this work and discuss the effect of Dzyaloshinskii--Moriya
interactions on the (short-range) valence bond physics of the kagom\'{e} lattice in
the context of the material $\text{ZnCu}_3\text{(OH)}_6\text{Cl}_2$.

We begin by assuming the material is well described, to leading order, by a spin 1/2
kagom\'{e} Heisenberg antiferromagnet, though the details of our analysis do not depend specifically
on the KHAF and will apply for any Hamiltonian with the assumed spectral properties of Fig.~\ref{fig:spec}.  We review some known properties of the KHAF which suggest that this model
satisfies the assumptions of our formalism.  We then discuss the results of the calculation of $\chi_\text{powder}$ in light of recent experiments.  We note that while the KHAF assumption underlies
most theoretical treatments of this material to date, there is not currently a consensus on what model
best describes herbertsmithite.\cite{Ofer08}

\subsection{Model}

As noted in the Introduction, exact diagonalization studies of the KHAF suggest that
the ground state is non-magnetic\cite{Leung93, Waldtmann98} with a spin gap of order $J/20$.\cite{Waldtmann98}
The gap is filled with an exponentially (in the system size) large number of non-magnetic states which appear to become a gapless continuum in the thermodynamic limit.  It was noted in Ref.~[\onlinecite{Waldtmann98}] that a gapless singlet continuum and spin gap an order of magnitude smaller than $J$ could indicate the importance of longer than nearest-neighbor valence bonds.  However, in Ref.~[\onlinecite{Mambrini00}], it was noted that at least the gapless singlet continuum, could be reproduced by considering the KHAF restricted to the nearest-neighbor valence bond (dimer) subspace.  Indeed, dimer states with gapless singlet excitations have been seen in toy models.\cite{Rokhsar88}  In addition, it was noted that the dimer subspace is the simplest subspace that captures the feature of an exponentially growing number of states.

More recently, the series expansion study of Ref.~[\onlinecite{Singh07a}] examined the energies of
various dimer coverings of the kagom\'{e} for the KHAF Hamiltonian  and found that a particular one, namely the ``perfect hexagon" state first noted in Ref.~[\onlinecite{Marston91}], was optimal though all dimer coverings were very close in energy, with a bandwidth (of order $J/50$) small compared to the
spin gap.  Their estimate for the ground state energy per site compared well with exact diagonalization 
studies of the model.  Put together, these facts suggest that spectra such as Fig.~\ref{fig:spec}ab, where the low energy physics is determined by a narrow band of dimer states, are decent caricatures of what happens in the KHAF.  Further arguments about the suitability of restricting attention to the dimer subspace when studying low energy properties of the KHAF may be found in section IIF of Ref.~[\onlinecite{Misguich03}].

In Ref.~[\onlinecite{Singh08a}], the elementary triplet (and singlet) excitations of the above mentioned ``perfect hexagon" valence bond crystal state were studied.  For that state, it was found that the lowest lying triplets have a spin gap of order $0.08 J$ but a bandwidth of only $\sim 0.01 J$.  This is consistent with the assumption of a narrow $S=1$ band required in section \ref{sec:model}.

Therefore, while currently there is no Hamiltonian on the kagom\'{e} lattice for which it can be \emph{explicitly} shown that the assumptions of section \ref{sec:model} are satisfied, the above facts suggest that these assumptions are sensible with respect to the KHAF and related models.

In adapting Eq.~(\ref{eq:chipowder3}) to the kagom\'{e} lattice, cases 1 and 5 (see Table \ref{tab:gen}) are straightforward (using $z$=4 for the kagom\'{e}) and case 9 does not occur.  For the kagome lattice, each dimer is part of a triangular plaquette and each plaquette contributes $(D^z)^2+(D^p)^2\cos \frac{2\pi}{3} = (D^z)^2 - \frac{1}{2}(D^p)^2$ to sum (2) in Eq.~(\ref{eq:chipowder2}).  Therefore,
\begin{equation}
\chi_\text{powder}\approx \frac{N(g\mu_B)^2}{4\Delta^3} \left (2D^2-(D^z)^2 \right ) + O(x^6).
\label{eq:kagchi}
\end{equation}

We can also use our formalism to calculate the anisotropy in $\chi$, as would be seen in
measurements on single crystals:
\begin{equation}
\frac{\chi^{z}}{\chi^{p}}=2\left (1-\left(\frac{D^z}{D}\right)^2\right)
\label{eq:kaganiso}
\end{equation}

\subsection{Comparison with experiment}

In comparing our results with experiments on $\text{ZnCu}_3\text{(OH)}_6\text{Cl}_2$ , the first point to note is that DM does not induce a zero field magnetization, which is consistent with the lack of magnetic order according to a number of techniques.\cite{Helton07,Ofer07,Mendels07,Imai08}

Eq.~(\ref{eq:kagchi}) is an expression for the $T=0$ powder susceptibility.  In order to compare with
experiment, we assume that $D^z\approx D$; $\Delta\sim {J}/{20}$~\cite{Waldtmann98}
or possible a bit higher $\sim {J}/{10}$~\cite{Singh08a} (the gap entering the calculation is the one separating the bands connected by $H_\text{DM}$ for which the spin gap is a lower bound); and $J\sim 170$~K.\cite{Rigol07a}

There are two sets of experiments which suggest that the susceptibility saturates in the $T\rightarrow 0$ limit.  Ofer et al.\ \cite{Ofer07} measured $\chi_\text{powder}$ using $\mu$SR in a 2~kG magnetic field in a temperature range from around 100~mK to around 200~K.  They observed a monotonic rise in $\chi_\text{powder}$ as the temperature was lowered and the last two data points indicated a saturation value of $\chi_\text{powder}(T=0)\sim 15.7\times 10^{-3}$ emu/mol Cu.  Using Eq.~(\ref{eq:kagchi}), this value
is consistent with $D$ ranging from 0.03~$J$ to 0.08~$J$.  However, other interpretations of this data
have involved the Ising anisotropy\cite{Ofer08} and/or stressed the role of impurities\cite{Bert07, Gregor08a,Chitra08, Rozenberg08}.

The second experiment is a recent $^{17}$O NMR study\cite{Olariu08} purported to measure the intrinsic susceptibility of the kagome planes.  Because the energy scale of the 6.5 $T$ field used in that study is comparable to our lower estimate for $\Delta$, it is not clear that Eq.~(\ref{eq:kagchi}), the derivation of which assumed that the Boltzmann factors of the excited bands were relatively small compared to the low energy dimer manifold, will directly apply.  However, if the actual gap is closer to the higher end of our range $\sim {J}/{10}$, then Eq.~(\ref{eq:kagchi}) might produce a reliable order of magnitude estimate.
The measurements of Ref.~[\onlinecite{Olariu08}] suggest\cite{Cepas08} a $T=0$ susceptibility per spin of $\chi=0.13$ which implies a molar susceptibility of $\sim 1.1\times 10^{-3}$ emu/mol Cu.  From Eq.~(\ref{eq:kagchi}), this implies that $D$ ranges from 0.008 $J$ to 0.02 $J$.

We caution against taking these estimates too literally because in both cases they are based on a relatively small number of data points and in both experiments, it is not clear that DM interactions are the only factors at play.  Also, because these two experiments suggest qualitatively different temperature
dependences for $\chi$, they can not both be measurements of the quantity we are calculating.  However, we would like to emphasize that these estimates strongly suggest that the magnitude of the DM interaction may be significantly smaller than previously quoted values.  We will discuss this point in more detail in the next section.

While our calculation is for the $T=0$ susceptibility, we can speculate on what happens
at a small but nonzero temperature.  Numerical studies of the KHAF\cite{Waldtmann98}
suggest that the singlet sector has a linear density of states at the very lowest energies.  An approximate way of accounting for this is for a low energy state with energy $\epsilon$ above the ground state, the gap entering Eq.~(\ref{eq:energypert4}) will be $\Delta-\epsilon$, if we assume the triplet band is still flat.  The sum over $n$ in Eq.~(\ref{eq:chipowder2}) will become an integral over a density of states, the range of integration being the effective low energy bandwidth $\delta$.  In the limit where $T\ll\delta\ll\Delta$, the result will be a modification of Eq.~(\ref{eq:kagchi}): $\chi(T)\approx\chi(T=0)(1+CT)$ where $C$ is a positive constant that depends on $D$ and $\Delta$.  Therefore, our picture predicts that $\chi$ should rise with $T$ so in this sense resembles what is seen in Ref.~\onlinecite{Olariu08}.  However, more experiments and a more refined theoretical treatment are clearly warranted.

\section{Discussion}
\label{sec:discuss}

The original motivation for this work was the material ZnCu$_3$(OH)$_6$Cl$_2$,
which appears to have a non-magnetic low temperature phase, whose nature has been the
subject of much speculation.  At the beginning of this paper, we noted that any
non-magnetic phase could be viewed as either a short-range or long-range valence bond
phase.  In the previous section, we have shown that a short-range valence bond phase with a very small DM coupling provides a picture of the low temperature phase of ZnCu$_3$(OH)$_6$Cl$_2$ which reconciles the lack of observed magnetic order, the lack of an observed spin gap,
and known facts about the KHAF.  We now discuss this picture in the light of more recent
experiments and other theories.

Recently, Zorko et al.\ \cite{Zorko08} determined the magnitudes of $D^z$ and $\mathbf{D}^{\text{p}}$ based on an analysis of high temperature ESR measurements.  They found the best fit was
obtained for $D^z\sim 0.08 J$ and $|\mathbf{D}^\text{p}| \sim 0.01 J$.  These values were somewhat smaller than the estimates of Rigol and Singh\cite{Rigol07a}, also based primarily on high temperature measurements.  One possible reason for our discrepancy with these estimates is the
observation of Imai et al.\cite{Imai08} that the OH bonds, which
mediate the superexchange between the Cu spins, freely rotate about the Cu-Cu axis at high
temperature but freeze in random orientations below 50 K.  While the effect of this on $J$ would
be weak, we expect the effect on $\mathbf{D}$ to be more significant because the direction of
$\mathbf{D}$ is determined by the position of this OH group.\footnote{This point was also made in
the discussion section of Ref.~[\onlinecite{Cepas08}].}  Therefore, there is strong reason to suspect
that the strength of the DM interaction experienced by the system at low $T$ is different than the value suggested both by the high temperature ESR analysis and the numerical fit of the high temperature magnetic susceptibility\cite{Rigol07a}.  We emphasize again that our estimate of $D$ is based on
low $T$ susceptibility data.  (However, it remains to be seen if an analysis of the low $T$ ESR data will
lead to a larger or smaller estimate for $D$.)

There is another possible reason for the discrepancy with Ref.~[\onlinecite{Zorko08}].  One could speculate that the ESR analysis of Ref.~[\onlinecite{Zorko08}] was based on the assumption that the symmetric anisotropic exchange could be neglected in comparison to the antisymmetric DM interaction.  While this is often a reasonable approach, given that the former term is quadratic in the spin-orbit coupling while the latter is linear, some recent magnetization measurements\cite{Ofer08} give reason to suspect that the symmetric term might also be important in this material. (The importance of accounting for such a term has been pointed out by Shekhtman et al.\cite{Shekhtman92,Shekhtman93} and has received further experimental confirmation\cite{Zheludev98}.) However, in the present context, we again do not know whether accounting for this will lead to a larger or smaller estimated value of $D$.

One reason to believe that our estimates may not be entirely off target is the fact that in the presence of sufficiently strong $D^z$, the KHAF is expected to order magnetically. For classical spins, the presence of such a coupling, no matter how small, would favor a long-range three-sublattice $120^{\circ}$ magnetic order in the $x-y$ plane (with the chirality determined by the sign of $D^z$)\cite{Elhajal02}. Curiously, a similar result would appear to hold in the quantum limit if one were to adopt a proposed algebraic spin liquid description of the KHAF.\cite{Ran07a,Hermele08}
However, no signs of any magnetic ordering in herbertsmithite have been observed thus far in contradiction with both the quasi-classical picture, even after accounting for the spin-wave corrections\cite{Ballou03}, and the algebraic spin liquid picture.

Cepas et al.\cite{Cepas08} recently studied the KHAF augmented with DM via exact diagonalization.  They found that the non-magnetic phase observed in the $D=0$ case\cite{Leung93,Waldtmann98} is stable to small DM coupling until at $D_c \sim 0.1 J$ the system undergoes a phase transition to a magnetically ordered phase. The value of $D^z$ obtained in Ref.~[\onlinecite{Zorko08}] is very close to such a magnetic transition while our range of estimates puts the system deeper into the non-magnetic phase, making it definitely consistent with the current lack of any experimentally observed magnetic order or phase transition.  Relevance to experiments aside, our estimate and Ref.~[\onlinecite{Cepas08}] together form a theoretically self-consistent picture because our calculation assumes a non-magnetic phase.

Assuming the picture of a short-range valence bond phase is correct, an obvious question would be what is the nature of the phase?  It is difficult to address this question using our formalism.  As mentioned previously, the overlap expansion will be more accurate, and hence our calculation more reliable, for a valence bond crystal  where the fluctuations are weak (such as the columnar state\cite{Rokhsar88} on the square lattice) as opposed to a liquid or a valence bond solid where fluctuations are strong (for example, a plaquette phase\cite{Moessner01c}).   Phases based on maximizing ``perfect hexagons'', variants of which were suggested by Marston and Zeng\cite{Marston91}, Nikoli\'{c} and Senthil\cite{Nikolic03}, and more recently by Singh and Huse\cite{Singh07a}, are candidate states to which our calculations might reliably apply.  We hope the present work will rekindle interest in these and other short-range valence bond phases in the context of herbertsmithite.

A non-magnetic alternative to the above picture is a phase based on long-range valence bonds.  As these phases are spin-gapless by construction, the absence of a spin gap in the experiment is explained by fiat.  One issue with this approach is the aforementioned numerical evidence suggesting the KHAF has a spin gap.  Another issue is that at least one calculation\cite{Ran07a} of the properties of such a phase, a variant of an algebraic spin liquid, shows the susceptibility vanishing at low temperatures as $\chi\sim T$, instead of saturating.  An already mentioned recent calculation\cite{Hermele08} indicates that DM interactions will drive this same phase into a magnetically ordered state, which has not been seen in experiment so far.  A nonzero susceptibility can, in principle, be obtained within a long-range valence bond picture by considering phases that break SU(2) invariance\cite{Ryu07}, by coupling the liquid phase to impurities\cite{Kolezhuk06, Ran07a}, or via a state with a spinon Fermi surface\cite{Motrunich04}.  

To conclude, we would like to reiterate or central result:  we have shown that Dzyaloshinskii--Moriya interactions can account for a nonzero value for the $T=0$ magnetic susceptibility in a system dominated by short-range valence bonds at low temperatures, while the magnetization remains zero.  Applied to herbertsmithite, we find that such a picture is consistent with experiments and our approach may be used to estimate the strength of the Dzyaloshinskii--Moriya coupling.  There are, however, inconsistencies in attempting to reconcile our results to the estimates based on the high-temperature data. Clearly, further studies, both theoretical and experimental are needed to fully clarify these issues.  Progress on this front would contribute greatly towards understanding this material.

\section{Acknowledgements}
We would like to express our gratitude to Michael Hermele for an important discussion on calculating the susceptibility in systems that do not conserve spin, which led to a serious revision of this paper.  The authors are also indebted to Roland Kawakami,  Douglas MacLaughlin, Roderich M\"{o}ssner, Marcelo Rozenberg, Rajiv Singh, Chandra Varma, and Ashvin Vishwanath for many valuable discussions and suggestions. We are also grateful to Amit Keren both for sharing his data and for providing useful feedback. In addition, KS would like to thank the Aspen Center for Physics for its hospitality. This research has been in part supported by the NSF under grant DMR-0748925.

\bibliographystyle{apsrev}
\bibliography{corr}

\end{document}